\documentclass[letterpaper]{jpconf}
\usepackage{verbatim}
\usepackage[dvips]{graphicx}\unitlength=1mm
\usepackage[hmargin=3cm,vmargin=3.5cm]{geometry}
\usepackage[square,sort&compress,numbers]{natbib}
\usepackage{hyperref}

\def\slashed#1{\kern+0.1em /\kern-0.55em #1}
\begin{document}
\title{Lattice QCD and Hydro/Cascade Model of Heavy Ion Collisions}

\author{Michael Cheng}
\ead{cheng24@llnl.gov}
\address{Lawrence Livermore Laboratory\\
		7000 East Avenue, L-415\\
		Livermore, CA 94550}

\begin{abstract}
We report here on a recent lattice study of the QCD transition region at finite temperature
and zero chemical potential using domain wall fermions (DWF).  We also present a 
parameterization of the QCD equation of state
obtained from lattice QCD that is suitable for use in hydrodynamics studies of heavy ion collisions.
Finally, we show preliminary results from a multi-stage hydrodynamics/hadron cascade
model of a heavy ion collision, in an attempt to understand how well the experimental data
(\textit{e.g.} particle spectra, elliptic flow, and HBT radii) can constrain the inputs 
(\textit{e.g.} initial temperature, freezeout temperature, shear viscosity, equation of state) of the 
theoretical model.
\end{abstract}

\section{Introduction}
The aim of the various high energy heavy ion collision (HIC) programs at experimental facilities
such as RHIC, SPS, LHC, and FAIR is to study the properties of nuclear matter under the
extreme conditions of high energy and high density.  In particular, at sufficiently high energy
density, it is predicted that normal hadronic matter will undergo a transition into a wholly new
state of matter, the quark-gluon plasma (QGP) \cite{Collins:1974ky, Shuryak:1980tp}, where the constituent quarks and 
gluons will no longer be confined within hadronic states and chiral symmetry is restored.

In principle, the properties of the QGP can be calculated directly from the underlying theory, 
Quantum Chromodynamics (QCD).  In practice, \textit{ab initio} calculations via lattice QCD
require very large-scale computing resources.  It is only recently that high-precision lattice
QCD results\cite{Aoki:2005vt, Aoki:2006br,Aoki:2009sc, Cheng:2006qk, Cheng:2007tp,Bazavov:2009zn}  have become available that have the potential to quantatively
constrain models of heavy ion collisions.  In Sec. \ref{sec:DWF} we discuss a lattice calculation\cite{Cheng:2009be} 
that uses the domain wall fermion method\cite{Kaplan:1992bt} to calculate
the crossover temperature of QCD at zero chemical potential.
In section \ref{sec:EoS} we present a parameterization of a high-precision lattice calculation
of the QCD Equation of State (EoS) that is useful for modeling heavy ion collisions.

In addition to theoretical calculations of the QGP, a robust understanding
of the dynamics of a heavy ion collision is needed in order to translate experimental results
into constraints on the physical properties of hot QCD matter.  Several approaches, such as
Boltzman transport, hydrodynamics, and hadronic cascade \cite{Molnar:2001ux, Teaney:2000cw, Bleicher:1999xi, Bass:1998ca} have 
been used to model heavy ion collisions.  However, the relevant physics changes so
drastically over the lifetime of a HIC that it is difficult to capture all the aspects of a
collision with a single model.  In Sec. \ref{sec:Model} we present preliminary results on
a multi-stage model of a HIC that incorporates Glauber initial conditions with
pre-equilibrium flow, 2-d viscous hydrodynamics, Cooper-Frye freezeout, and a hadronic
cascade.

\section{QCD Transition using domain wall fermions}
\label{sec:DWF}

The location of the QCD crossover has been a subject of much debate in the past several
years \cite{Aoki:2006br, Aoki:2009sc, Cheng:2006qk}.  Because of computational expediency, many of these high-precision
studies have been done using the staggered fermion formulation.  Although computationally
inexpensive, staggered fermions have the known flaw that they only preserve a $U(1)$
subgroup of the full $SU(N_f)$ chiral symmetry at finite lattice spacing.  This manifests itself
in large lattice artifacts in certain quantities, \textit{e.g.} non-degeneracy in the pion spectrum.
Domain Wall Fermions (DWF) preserve the full $SU(N_f)$ chiral symmetry at finite lattice
spacing \cite{Kaplan:1992bt}. Thus, a DWF study at finite temperature is useful to test the robustness of the 
recent staggered studies of $T_c$ and helps us to better understand the role of chiral symmetry 
in the QCD crossover.

\begin{figure}[t]
\begin{center}
\includegraphics[width=0.60\textwidth]{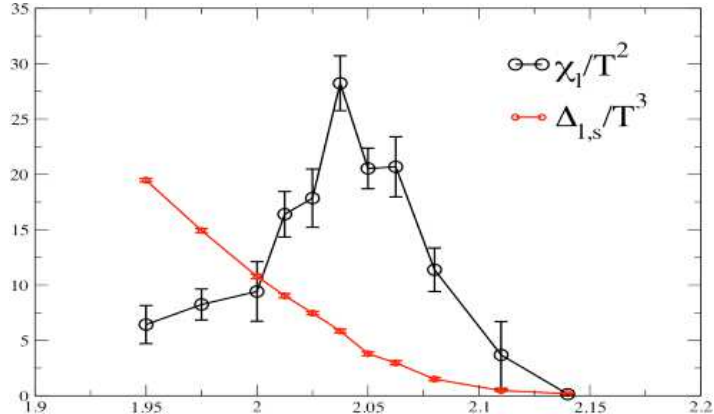}
\caption{The disconnected chiral susceptibility, $\chi_l/T^2$ and the subtracted chiral
condensate, $\Delta_{l,s}/T^3$. as a function of the bare lattice coupling $\beta = 6/g_0^2$.  The peak in the disconnected chiral susceptibility corresponds to $T \approx 170~ \textrm{MeV}$.}
\label{fig:dwfLs32}
\end{center}
\end{figure}

In our DWF calculation \cite{Cheng:2009be}, we perform simulations at 11 different temperatures in the 
range
$120 ~\textrm{MeV} < T < 220 ~\textrm{MeV}$ at zero chemical potential.  We use a temporal
lattice size of $N_t = 8$, which is related to the temperature via $T = 1/(N_t a)$.  These 
temperatures correspond to lattice spacings in the range $0.11 ~\textrm{fm} < a < 0.21 ~\textrm{fm}$.

Various quantities are used to locate the chiral-symmetry restoring and deconfinement
transitions.  For the chiral transition, we use the subtracted chiral condensate,
\begin{equation}
\frac{\Delta_{l,s}(T)}{T^3} = \frac{\left<\bar{\psi}_l\psi_l\right>  - \frac{m_l}{m_s} \left<\bar{\psi_s}\psi_s\right>}{T^3},
\end{equation}
where the subscripts $l,s$ denote the light and strange quarks, respectively. In the phase where
chiral symmetry is restored ($T > T_c$), we expect $\Delta_{l,s}(T)$ to vanish, while for
$T < T_c$, it is non-zero.  In addition to the chiral condensate, we also calculate the
disconnected chiral susceptibility,
\begin{equation}
\frac{\chi_l}{T^2} = VT^3 \left(\left<\left(\bar{\psi_l}\psi_l\right)^2\right> - \left<\bar{\psi_l}\psi_l\right>^2\right).
\end{equation}

\begin{figure}[t]
\vspace{-1cm}
\begin{minipage}{0.47\textwidth}
\begin{center}
\includegraphics[width=\textwidth]{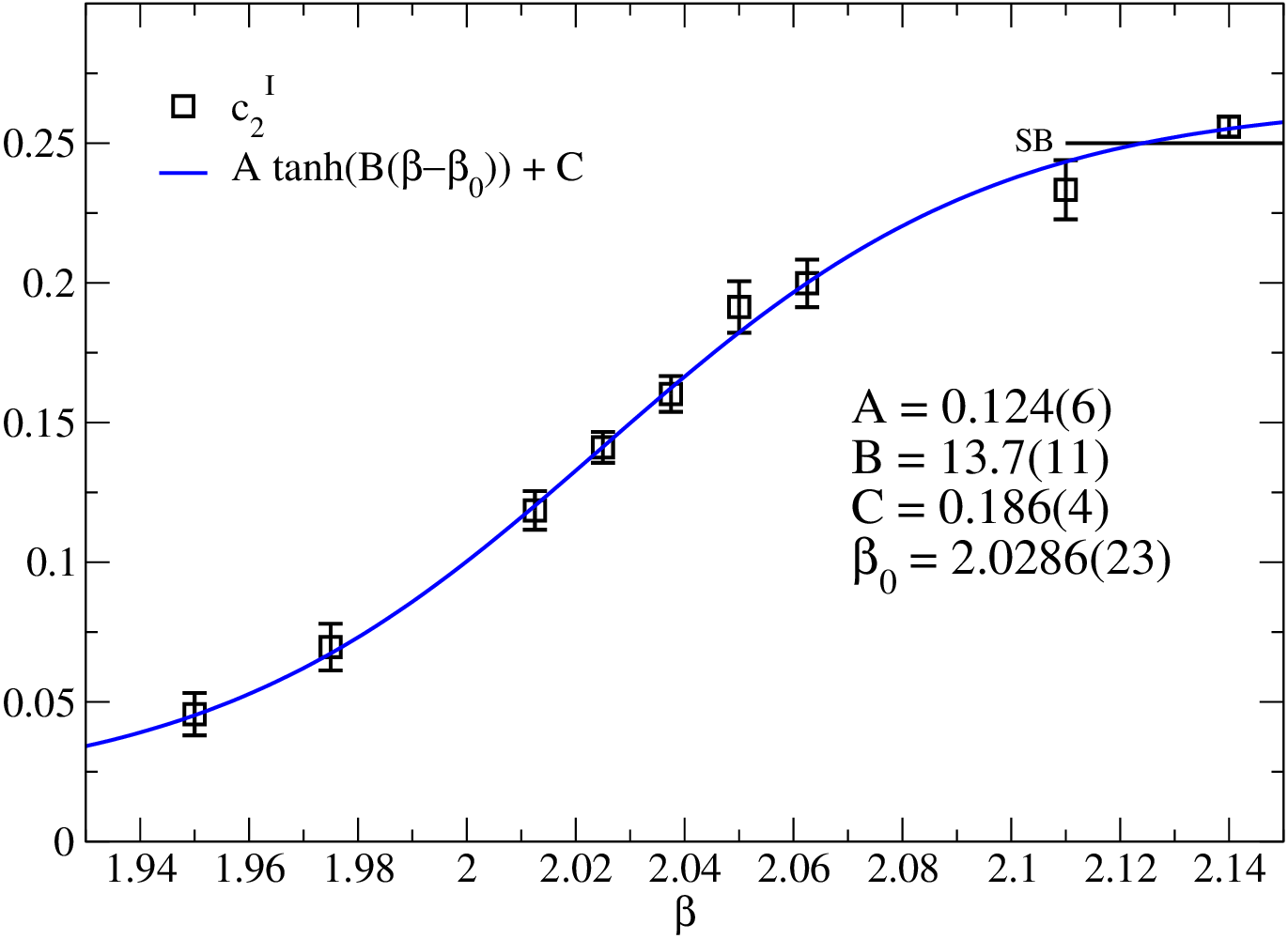}
\caption{The isospin susceptibility, $c_2^{I}$ versus bare coupling $\beta = 6/g_0^2$.}
\label{fig:c2I}
\end{center}
\end{minipage}
\hspace{0.05\textwidth}
\begin{minipage}{0.47\textwidth}
\begin{center}
\includegraphics[width=\textwidth]{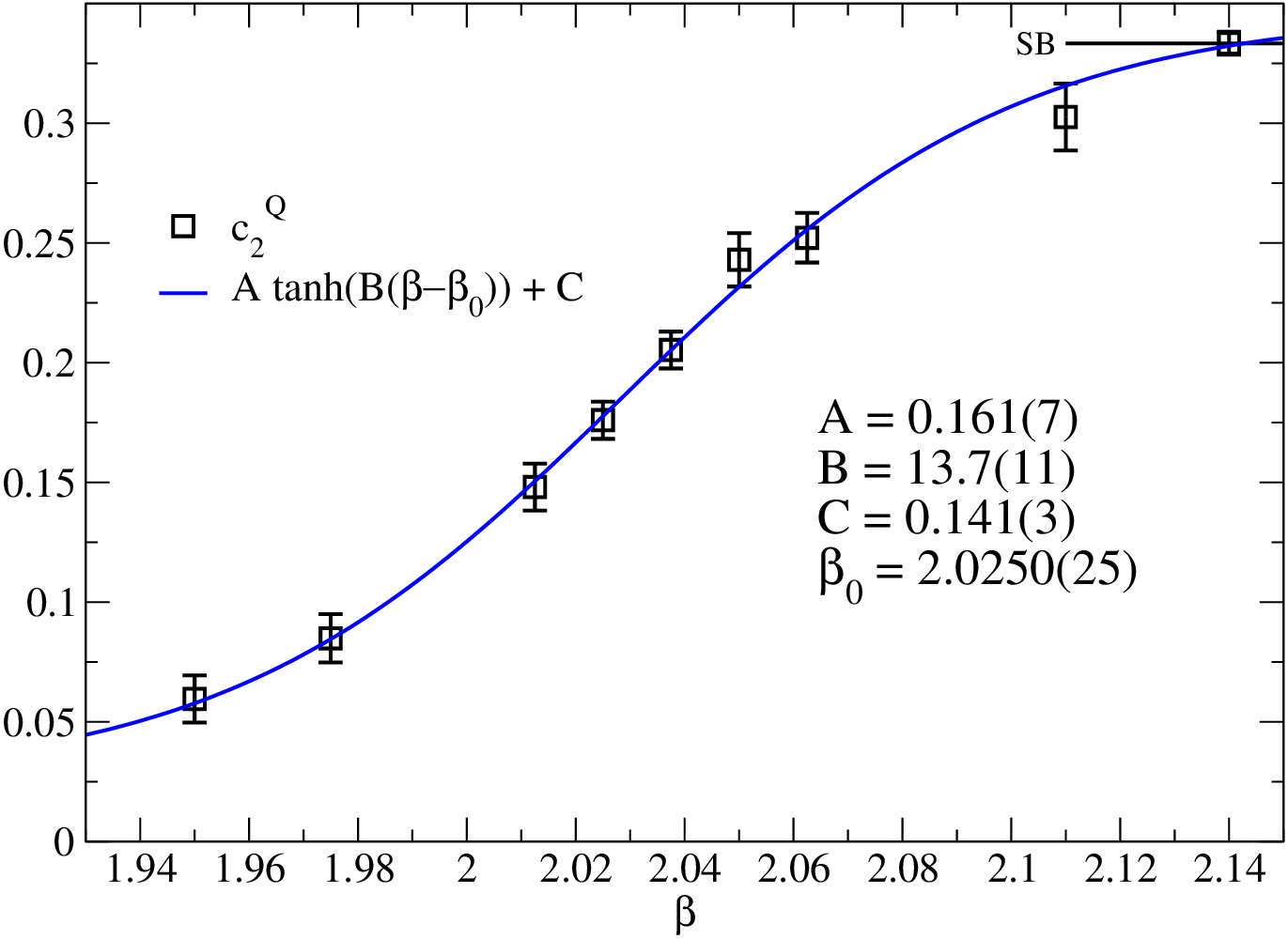}
\caption{The electric charge susceptibility, $c_2^{Q}$ versus bare coupling $\beta = 6/g_0^2$.}
\label{fig:c2Q}
\end{center}
\end{minipage}
\end{figure}

This quantity should exhibit a peak at $T = T_c$.  Fig. \ref{fig:dwfLs32} shows both $\frac{\Delta_{l,s}}{T^3}$ and $\frac{\chi_l}{T^2}$.  As can be seen, the disconnected
chiral susceptibility has a clear peak near $\beta = 6/g_0^2 = 2.03$, where $g_0$ is the
bare coupling on the lattice.  This corresponds to $T \approx 170$ MeV.

For the deconfinement transition, we calculate the electric charge susceptibility ($c_2^Q$) and
the isospin susceptibility ($c_2^I$):
\begin{equation}
\frac{c_2^X}{T^2} = \frac{1}{2}\frac{1}{VT^3}\frac{\partial^2 \ln Z}{\partial (\mu_X/T)^2}\vert_{\mu_X = 0};~\textrm{X = Q, I},
\end{equation}
where $\mu_Q$ and $\mu_I$ are the electric charge and isospin chemical potentials and
$\ln Z$ is the QCD partition function.  As can
be seen in Figs. \ref{fig:c2I} and \ref{fig:c2Q}, these deconfinement observables  show a rapid rise 
in the same general region as those observables related to chiral symmetry.

Although the chiral susceptibility has a clear peak near $T \approx 170~ \textrm{MeV}$ and
the electric charge and isospin susceptibilities imply $T_c$ in the same region, our calculation
still suffers from significant uncertainties.  The spatial volume used is quite small, from $V = (1.7 ~
\textrm{fm})^3$ up to $V = (3.1 ~\textrm{fm})^3$, so our result may be polluted with uncontrolled
finite-volume effects.  In addition, the light quark masses are not physical, but correspond
to $m_\pi = 300$ MeV near the susceptibility peak.  Furthermore, the simulations are not
done on a line of constant quark mass, so that $m_\pi$ is effectively much heavier at low 
temperatures than at high temperatures.  These effects introduce significant uncertainty into our 
estimate of $T_c$, so we are only able to quote a wide range of possible values: 
$150 ~\textrm{MeV} < T < 190~ \textrm{MeV}$.  
Unfortunately, this does not do much to discriminate between the results of recent lattice 
calculations, but we hope to reduce these uncertainties in future calculations.

\begin{figure}[t]
\begin{center}
\vspace{-1cm}
\includegraphics[width=0.60\textwidth]{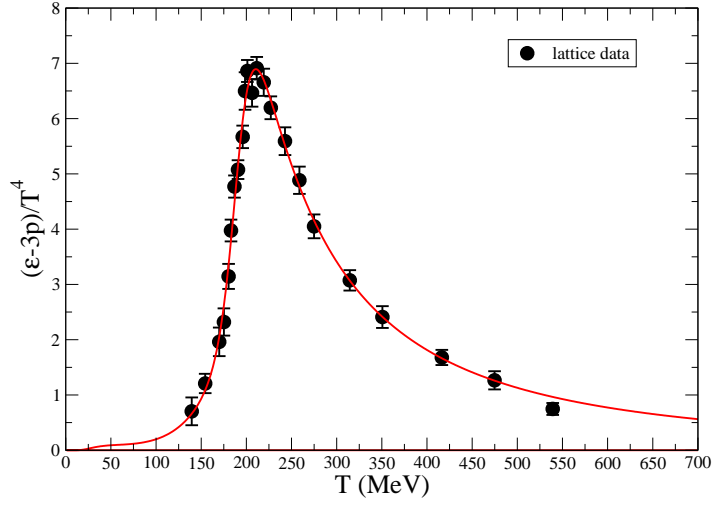}
\caption{Lattice QCD results of the interaction measure $(\epsilon - 3p)/T^4$ with corresponding
fit to the parameterization $I(T)= (1/I_{low}(T) + 1/I_{high}(T))^{-1}$.}
\label{fig:p4fit}
\end{center}
\end{figure}

\section{Parameterization of the lattice QCD equation of state}
\label{sec:EoS}

There have been several attempts to parameterize the lattice results for the QCD equation of state
(EoS) in terms of a continuous function that can be conveniently input into hydrodynamic models.  
Our parameterization reproduces the expected asymptotic
behavior of the QCD equation of state in the limits $T\rightarrow 0$ and $T \rightarrow \infty$,
smoothly joining them together in the intermediate regime near the QCD crossover.

In lattice calculations, the interaction measure (I(T) = $\epsilon - 3p$) is the thermodynamic 
quantity that  is naturally calculated, from which all other thermodynamics quantities such as the 
pressure $p$, energy density $\epsilon$, entropy density $s$, and speed of sound $c_s$ can be 
derived.  Thus, it is sufficient to parameterize the interaction measure.

In the high temperature limit, the interaction measure is expected to have the form \cite{Pisarski:2006yk,Megias:2007pq}:
\begin{equation}
\frac{I_{high}(T)}{T^4} = \frac{\alpha_1}{T^4} + \frac{\alpha_2}{T^4}.
\end{equation}
The lattice QCD data agrees well with this expected behavior.  At low temperatures, QCD can
be described as a gas of mesons and baryons.  A widely-used approximation is to assume
that this gas is non-interacting.  This is called the Hadron Resonance Gas (HRG) model \cite{Hagedorn:1965st}.

The existing lattice results give an interaction measure that tends to be lower than that given by
the HRG model at low temperatures. Thus, in our parameterization, deviations from
the HRG model at low temperature are included in order to accurately fit the lattice data:
\begin{equation}
\frac{I_{low}(T)}{T^4} = \frac{I_{HRG}(T)}{T^4} \left(\alpha_3 +\left( \frac{T}{\alpha_4}\right)^{\alpha_5}\right).
\end{equation}

\begin{figure}[t]
\begin{center}
\vspace{-1cm}
\includegraphics[width=0.60\textwidth]{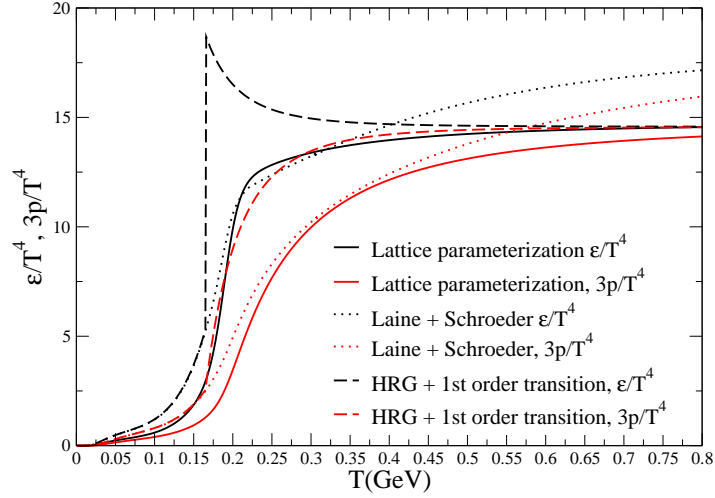}
\caption{Comparison of the energy density, $\epsilon$ and three times the pressure, 3p, for
our lattice parameterization, the EoS from Laine and Schroeder 
\cite{Laine:2006cp}, and the Hadron Resonance Gas Model with a 1st order transition.}
\label{fig:eos_comp}
\end{center}
\end{figure}

\begin{figure}[hbt]
\begin{center}
\vspace{0.5cm}
\includegraphics[width=0.60\textwidth]{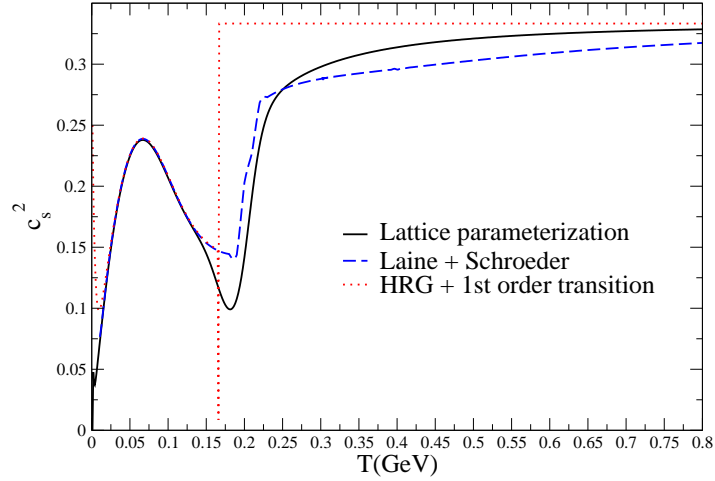}
\caption{Comparison of the speed of sound squared, $c_s^2$, for
our lattice parameterization, the EoS from Laine and Schroeder
\cite{Laine:2006cp}, and the Hadron Resonance Gas Model with a 1st order transition.}
\label{fig:cs2_comp}
\end{center}
\end{figure}

The low and high temperature parameterizations are combined via:
\begin{equation}
I(T) = \left(\frac{1}{I_{low}(T)} + \frac{1}{I_{high}(T)}\right)^{-1}
\end{equation}
and the parameters ($\alpha_1 \cdots \alpha_5$) can be extracted through a fit to the
lattice data.  Note that fixing $\alpha_3 = 1$ is also an option, if one wants to
force the parameterization onto the HRG result as $T \rightarrow 0$.

Table \ref{tab:p4fit} tabulates the fit parameters ($\alpha_1 \cdots \alpha_5$) from a fit to the
lattice EoS for the p4 action at $N_t = 8$ \cite{Bazavov:2009zn}.  Fig. \ref{fig:p4fit} shows the result
of this parameterization plotted along with the actual lattice data.  Fig. \ref{fig:eos_comp} shows 
a comparison of this lattice parameterization with two other forms for the equation of state 
often used in hydrodynamic modeling - a form proposed by Laine and Schroeder \cite{Laine:2006cp}, 
and one where the HRG EoS at low temperature is joined to an ideal gas at high temperature with 
a first-order phase transition at $T_c = 165$ MeV.  Figure \ref{fig:cs2_comp} shows the same 
comparison  for the speed of sound squared, $c_s^2$.

\begin{table}[hbt]
\begin{center}
\begin{tabular}{c|c|c|c|c}
$\alpha_1$ & $\alpha_2$ & $\alpha_3$ & $\alpha_4$ & $\alpha_5$\\
\hline
0.26867 & 0.00345 & 0.50120 & 0.18529 & 15.17516
\end{tabular}
\caption{Fit parameters for $I(T)$.}
\label{tab:p4fit}
\end{center}
\end{table}

\section{Hydro+Cascade Model}
\label{sec:Model}
We introduce a multi-stage model where a heavy ion collision is modeled from the first moment
of collision until the final state hadrons are effectively free-streaming to the particle detectors.
In constructing this model, we wish to systematically test the effect of varying various model
inputs and to see which set of parameters give the best agreement with experimental results.
Among the things that we wish to test are: the nature of the initial conditions, the inclusion of
pre-equilibirum flow, the initial temperature $T_i$, the shear viscosity $\eta$, the freezeout
temperature $T_f$, and the equation of state.

\subsection{Initial Conditions}
For initial conditions, we use an optical Glauber model \cite{Glauber:1955qq,Miller:2007ri}, where the initial energy density is 
proportional to the number of participant nuclei, $N_{part}$ or the number
of binary collisions, $N_{coll}$.  The magnitude of the
energy density is chosen so that the maximum energy density $\epsilon_{max}$ corresponds
to some initial temperature, $T_i$.  The impact parameter, $b$ can
be directly varied to obtain different centralities.

A commonly used assumption
is that the initial flow velocities vanishes, neglecting the fact that flow velocities may develop
in the first $\tau \approx 1$ fm/c, before hydrodynamic evolution is applicable \cite{Huovinen:2006jp}.  Recently,
it has been found that neglecting initial flow velocities might have significant effects on 
collective observables, particularly those related to the source size \cite{Pratt:2008qv, Vredevoogd:2008id}.  Therefore,
we allow the modification of our initial conditions to account for possible pre-thermalization
flow.

\begin{figure}[t]
\vspace{-1cm}
\begin{minipage}{0.42\textwidth}
\begin{center}
\includegraphics[width=\textwidth,angle=-90]{figs/combined_yield.epsi}
\caption{$\pi^+$ invariant yield.  Experimental data from PHENIX results in the $10-15\%$ centrality bin \cite{Adler:2003cb}.  Model results obtained with impact parameter $b = 4.4 \textrm{fm}$, $T_i = 300 \textrm{MeV}$, $T_f = 150 \textrm{MeV}$, without initial flow and using the "lattice-inspired" EoS.}
\label{fig:yield}
\end{center}
\end{minipage}
\hspace{0.15\textwidth}
\begin{minipage}{0.42\textwidth}
\begin{center}
\includegraphics[width=\textwidth,angle=-90]{figs/combined_v2.epsi}
\caption{$\pi^+$ elliptic flow ($v_2$).  Experimental data from PHENIX results in the $0-10\%$ centrality bin \cite{Afanasiev:2009wq}.  Model results obtained with impact parameter $b = 4.4 \textrm{fm}$, $T_i = 300 \textrm{MeV}$, $T_f = 150 \textrm{MeV}$, without initial flow and using the "lattice-inspired" EoS.}
\label{fig:v2}
\end{center}
\end{minipage}
\end{figure}

\subsection{Viscous Hydrodynamics}
Because the relativistic generalizations of the Navier-Stokes equation are acausal, it is not
entirely clear how to formulate relativistic dissipative hydrodynamics.  One attempt is
the Israel-Stewart formalism \cite{Israel:1979wp}, where a relaxation time is introduced for every transport  coefficient.
However, there is still ambiguity as to which higher order terms to include.  In the case of
vanishing bulk viscosity, however, it has been shown \cite{Baier:2007ix} that the most general form includes
five additional terms, $\tau_\Pi, \kappa, \lambda_1, \lambda_2, \lambda_3$, where $\kappa = 0$
in flat space-time.  This formulation was implemented by Romatshke \cite{Luzum:2008cw}, and is the
one that we utilize.

Romatschke's code \cite{Luzum:2008cw} is two-dimensional, where the bulk viscosity ($\zeta$) 
implicitly vanishes, but the shear viscosity ($\eta$) can be non-zero.  For the relaxation times, we 
set $\tau_\Pi = 6 \eta/s T$, $\lambda_1 = 0$, $\lambda_2 = -2 \eta/\tau_\Pi$, $\lambda_3 = 0$,
which are valid choices for a weakly-coupled plasma.
The start time for the hydrodynamics evolution is taken to be $\tau_0 = 1 fm/c$, with the
initial conditions discussed above.

For the equation of state, we plan to test four different variations: 1) A HRG EoS for $T < T_c$ with 
a first order transition, 2) the "Laine-Schroder" EoS \cite{Laine:2006cp} used
in \cite{Luzum:2008cw}, 3) a parameterization of the lattice QCD EoS discussed in Sec. \ref{sec:EoS}, and
 4) the parmeterization of the lattice QCD EoS with $\alpha_3 = 1$ so that it is smoothly joined to 
 the HRG EoS at low temperatures.  A comparison of the first three of these can be seen in Figs. 
\ref{fig:eos_comp} and \ref{fig:cs2_comp}.

\begin{figure}[t]
\vspace{-1cm}
\begin{minipage}{0.42\textwidth}
\begin{center}
\includegraphics[width=\textwidth,angle=-90]{figs/combined_Rout.epsi}
\caption{$\pi^+$ $R_{out}$.  Experimental data from PHENIX results in the $0-30\%$ centrality bin \cite{Adler:2004rq}.  Model results obtained with impact parameter $b = 4.4 \textrm{fm}$, $T_i = 300 \textrm{MeV}$, $T_f = 150 \textrm{MeV}$, without initial flow and using the EoS from Laine and Schroeder.}
\label{fig:Rout}
\end{center}
\end{minipage}
\hspace{0.15\textwidth}
\begin{minipage}{0.42\textwidth}
\begin{center}
\includegraphics[width=\textwidth,angle=-90]{figs/combined_Rside.epsi}
\caption{$\pi^+$ $R_{side}$.  Experimental data from PHENIX results in the $0-30\%$ centrality bin \cite{Adler:2004rq}.  Model results obtained with impact parameter $b = 4.4 \textrm{fm}$, $T_i = 300 \textrm{MeV}$, $T_f = 150 \textrm{MeV}$, without initial flow and using the EoS from Laine and Schroeder.}
\label{fig:Rside}
\end{center}
\end{minipage}
\end{figure}

\subsection{Cooper-Frye Freezeout}
In modeling the freezeout of hadrons from the QGP, we use sudden freezeout and the Cooper-
Frye prescription.  In this method, the hadrons are frozen out on a hypersurface in $(x,y,z,t)$ of
constant temperature $T = T_f$, or equivalently constant energy density.  This freezeout 
temperature can be varied to study the effects of early or late freezeout.  The single
particle spectrum using the Cooper-Frye method is given by:
\begin{equation}
E\frac{d N}{d^3 \bf{p}} = \frac{d}{(2 \pi)^3}\int p_\mu d\Sigma^\mu f(x^\mu,p^\mu),
\end{equation}
where $d$ is the degeneracy factor, $d\Sigma^\mu$ represents the normal to the freezeout
hypersurface, and $f(x^\mu,p^\mu)$ is the phase space density with non-equilibrium corrections
\cite{Pratt:2010jt}.

Because the hydrodynamic evolution occurs only in two spatial dimensions, we assume boost-
invariance along the longitudinal direction to produce the full freezeout spectrum for the
particles.

\subsection{Hadron Cascade}
After freezeout, one must take into account final state interactions and feed-down decays in 
order to extract the final particle spectra.  In order to do this, we take the particles produced
at freezeout and evolve them through a hadronic cascade.  The code that we choose is
the Ultrarelativistic Quantum Molecular Dynamics (UrQMD) code \cite{Bleicher:1999xi, Bass:1998ca}, a microscopic transport
code that explicitly takes into account particle decays and hadron-hadron collisions.

\subsection{Preliminary Results}
We have preliminary results for several model runs corresponding to Au+Au collisions
at $\sqrt{s} = 200$ GeV per nucleon pair.  The parameters that we have used are summarized
in Tab. \ref{tab:Model_params}.  However, we have not yet explored all possible variations
of the model inputs.  The impact parameter of $b = 4.4$ fm. was chosen to match
as closely as possible to the mid-centraility bins for PHENIX results.

To match to experimental data, we concentrate on three classes of observables:
1) the invariant yield $(1/ 2 \pi p_T) (d N/d p_T dy)$ at mid-rapidity, 2) the elliptic flow
$v_2$, and 3) the HBT radii, $R_{out}, R_{side}$, and $R_{long}$.
To determine the degree of agreement with experimental, we perform a joint fit of both the model
and experimental results to a common smoothing function.  For $v_2$ and the HBT radii, the
smoothing function we use is 
a set of orthonormal polynomials, the Chebyshev polynomials.  For the invariant yield, we
found that the product of Chebyshev polynomials with $\exp\left(-m_T/T\right)$, where
$m_T = m^2 + p_T^2$, produced the best fits.  Fig. \ref{fig:yield} shows the fit of the invariant
yield for $\pi^+$ with one set of model parameters.  Fig. \ref{fig:v2} shows the fit for
$v_2$ and Figs. \ref{fig:Rout} and \ref{fig:Rside} show HBT radii.

For each set of model parameters, the joint fit with the experimental data produces a value of 
$\chi^2$, which is a loose measure of the difference between the model and experiment.  Figs. \ref{fig:chi2_yield} and \ref{fig:chi2_v2} show the $\chi^2$ distributions for $\pi^+$
for the invariant yield and $v_2$ as a function of $T_i$ and $\eta/s$.

As this is a work in progress, we plan to perform a systematic exploration of the model
parameter space, and perform these joint fits with a larger set of experimental results.
We intend to understand how sensitive the final results are to the various model inputs, and
which model inputs give the best match to experimental results for various systems, energies,
and centralities.

\begin{figure}[t]
\vspace{-1cm}
\begin{minipage}{0.42\textwidth}
\begin{center}
\includegraphics[width=\textwidth,angle=-90]{figs/combined_yield_chi2_vs_T_etaos.epsi}
\caption{$\chi^2$ distribution for model/data fits in the $\pi^+$ invariant yield as a function of $T_i$ and $\eta/s$.}
\label{fig:chi2_yield}
\end{center}
\end{minipage}
\hspace{0.15\textwidth}
\begin{minipage}{0.42\textwidth}
\begin{center}
\includegraphics[width=\textwidth,angle=-90]{figs/combined_v2_chi2_vs_T_etaos.epsi}
\caption{$\chi^2$ distribution for model/data fits in $\pi^+$ $v_2$ as a function of $T_i$ and $\eta/s$.}
\label{fig:chi2_v2}
\end{center}
\end{minipage}
\end{figure}

\begin{table}[h]
\begin{center}
\begin{tabular}{c|c|c|c|c|c|c}
b (fm) & $N_{part}$ & $T_i$  (MeV)& $\eta/s$ & $T_f$ (MeV) & Initial Flow & EoS\\
\hline
4.4 & 270 & 250, 300, 350 & 0.08, 0.16, 0.24 & 150 & yes, no & Laine \& Schroder
\end{tabular}
\caption{A summation of the model parameters that have been explored so far for Au+Au
collisons at $\sqrt{s} = 200$ GeV per nucleon pair.}
\label{tab:Model_params}
\end{center}
\end{table}

\section*{Acknowledgements}
The work in Sec. \ref{sec:DWF} was done in collaboration with the RBC-Bielefeld Collaboration.
We thank K. Rajagopal for suggesting the parameterization we use in Sec. \ref{sec:EoS}, and
R. Soltz for producing the fits.  The work in Sec. \ref{sec:Model} is done in collaboration with D. 
Brown, I. Garishvili, A. Glenn, J. Newby, 
S. Pratt, and R. Soltz.  We thank P. Romatshke for providing his hydrodynamics code, and
the UrQMD collaboration for providing the code for the hadron cascade.  We also thank
P. Huovinen and D. Molnar for useful discussions.  Finally, we would in
particular like to thank the organizers of the Winter Workshop on Nuclear Dynamics 2010, R. Lacey
and S. Pratt, for allowing us to present this work.  This work performed under the auspices of the U.S. Department of Energy by Lawrence Livermore National Laboratory under Contract DE-AC52-07NA27344.

\bibliography{wwnd}

\providecommand{\newblock}{}
\begin{thebibliography}{10}
\expandafter\ifx\csname url\endcsname\relax
  \def\url#1{{\tt #1}}\fi
\expandafter\ifx\csname urlprefix\endcsname\relax\def\urlprefix{URL }\fi
\providecommand{\eprint}[2][]{\url{#2}}

\bibitem{Collins:1974ky}
Collins J~C and Perry M~J 1975 {\em Phys. Rev. Lett.\/} {\bf 34} 1353

\bibitem{Shuryak:1980tp}
Shuryak E~V 1980 {\em Phys. Rept.\/} {\bf 61} 71--158

\bibitem{Aoki:2005vt}
Aoki Y, Fodor Z, Katz S~D and Szabo K~K 2006 {\em JHEP\/} {\bf 01} 089
  (\textit{Preprint} \eprint{hep-lat/0510084})

\bibitem{Aoki:2006br}
Aoki Y, ~ Z, Katz S~D and Szabo K~K 2006 {\em Phys. Lett.\/} {\bf B643} 46--54
  (\textit{Preprint} \eprint{hep-lat/0609068})

\bibitem{Aoki:2009sc}
Aoki Y {\em et~al.\/} 2009 {\em JHEP\/} {\bf 06} 088 (\textit{Preprint}
  \eprint{0903.4155})

\bibitem{Cheng:2006qk}
Cheng M {\em et~al.\/} 2006 {\em Phys. Rev.\/} {\bf D74} 054507
  (\textit{Preprint} \eprint{hep-lat/0608013})

\bibitem{Cheng:2007tp}
Cheng M {\em et~al.\/} 2008 {\em Phys. Rev.\/} {\bf D77} 014511
  (\textit{Preprint} \eprint{0710.0354})

\bibitem{Bazavov:2009zn}
Bazavov A {\em et~al.\/} 2009 {\em Phys. Rev.\/} {\bf D80} 014504
  (\textit{Preprint} \eprint{0903.4379})

\bibitem{Cheng:2009be}
Cheng M {\em et~al.\/} 2010 {\em Phys. Rev.\/} {\bf D81} 054510
  (\textit{Preprint} \eprint{0911.3450})

\bibitem{Kaplan:1992bt}
Kaplan D~B 1992 {\em Phys. Lett.\/} {\bf B288} 342--347 (\textit{Preprint}
  \eprint{hep-lat/9206013})

\bibitem{Molnar:2001ux}
Molnar D and Gyulassy M 2002 {\em Nucl. Phys.\/} {\bf A697} 495--520
  (\textit{Preprint} \eprint{nucl-th/0104073})

\bibitem{Teaney:2000cw}
Teaney D, Lauret J and Shuryak E~V 2001 {\em Phys. Rev. Lett.\/} {\bf 86}
  4783--4786 (\textit{Preprint} \eprint{nucl-th/0011058})

\bibitem{Bleicher:1999xi}
Bleicher M {\em et~al.\/} 1999 {\em J. Phys.\/} {\bf G25} 1859--1896
  (\textit{Preprint} \eprint{hep-ph/9909407})

\bibitem{Bass:1998ca}
Bass S~A {\em et~al.\/} 1998 {\em Prog. Part. Nucl. Phys.\/} {\bf 41} 255--369
  (\textit{Preprint} \eprint{nucl-th/9803035})

\bibitem{Pisarski:2006yk}
Pisarski R~D 2007 {\em Prog. Theor. Phys. Suppl.\/} {\bf 168} 276--284
  (\textit{Preprint} \eprint{hep-ph/0612191})

\bibitem{Megias:2007pq}
Megias E, Ruiz~Arriola E and Salcedo L~L 2007 {\em Phys. Rev.\/} {\bf D75}
  105019 (\textit{Preprint} \eprint{hep-ph/0702055})

\bibitem{Hagedorn:1965st}
Hagedorn R 1965 {\em Nuovo Cim. Suppl.\/} {\bf 3} 147--186

\bibitem{Laine:2006cp}
Laine M and Schroder Y 2006 {\em Phys. Rev.\/} {\bf D73} 085009
  (\textit{Preprint} \eprint{hep-ph/0603048})

\bibitem{Glauber:1955qq}
Glauber R~J 1955 {\em Phys. Rev.\/} {\bf 100} 242--248

\bibitem{Miller:2007ri}
Miller M~L, Reygers K, Sanders S~J and Steinberg P 2007 {\em Ann. Rev. Nucl.
  Part. Sci.\/} {\bf 57} 205--243 (\textit{Preprint} \eprint{nucl-ex/0701025})

\bibitem{Huovinen:2006jp}
Huovinen P and Ruuskanen P~V 2006 {\em Ann. Rev. Nucl. Part. Sci.\/} {\bf 56}
  163--206 (\textit{Preprint} \eprint{nucl-th/0605008})

\bibitem{Pratt:2008qv}
Pratt S 2009 {\em Phys. Rev. Lett.\/} {\bf 102} 232301 (\textit{Preprint}
  \eprint{0811.3363})

\bibitem{Vredevoogd:2008id}
Vredevoogd J and Pratt S 2009 {\em Phys. Rev.\/} {\bf C79} 044915
  (\textit{Preprint} \eprint{0810.4325})

\bibitem{Adler:2003cb}
Adler S~S {\em et~al.\/} (PHENIX) 2004 {\em Phys. Rev.\/} {\bf C69} 034909
  (\textit{Preprint} \eprint{nucl-ex/0307022})

\bibitem{Afanasiev:2009wq}
Afanasiev S {\em et~al.\/} (PHENIX) 2009 {\em Phys. Rev.\/} {\bf C80} 024909
  (\textit{Preprint} \eprint{0905.1070})

\bibitem{Israel:1979wp}
Israel W and Stewart J~M 1979 {\em Ann. Phys.\/} {\bf 118} 341--372

\bibitem{Baier:2007ix}
Baier R, Romatschke P, Son D~T, Starinets A~O and Stephanov M~A 2008 {\em
  JHEP\/} {\bf 04} 100 (\textit{Preprint} \eprint{0712.2451})

\bibitem{Luzum:2008cw}
Luzum M and Romatschke P 2008 {\em Phys. Rev.\/} {\bf C78} 034915
  (\textit{Preprint} \eprint{0804.4015})

\bibitem{Adler:2004rq}
Adler S~S {\em et~al.\/} (PHENIX) 2004 {\em Phys. Rev. Lett.\/} {\bf 93} 152302
  (\textit{Preprint} \eprint{nucl-ex/0401003})

\bibitem{Pratt:2010jt}
Pratt S and Torrieri G 2010  (\textit{Preprint} \eprint{1003.0413})

\end{thebibliography}

\end{document}